\documentclass[aps,prl,
,reprint,
superscriptaddress,
showpacs,preprintnumbers,
 amsmath,amssymb,
 aps,
]{revtex4-1}
\usepackage{graphicx}
\usepackage{dcolumn}
\usepackage{bm}
\usepackage{hyperref}

\begin{document}

\preprint{Preprint}

\title{Precision Measurement for Metastable Helium Atoms of the 413~nm  Tune-Out Wavelength at Which the Atomic Polarizability Vanishes}

\author{B. M. Henson}
\affiliation{Research School of Physics and Engineering, Australian National University,\\ Canberra, Australian Capital Territory 0200, Australia}
\author{R. I. Khakimov}
\affiliation{Research School of Physics and Engineering, Australian National University,\\ Canberra, Australian Capital Territory 0200, Australia}
\author{R. G. Dall}
\affiliation{Research School of Physics and Engineering, Australian National University,\\ Canberra, Australian Capital Territory 0200, Australia}
\author{K. G. H. Baldwin}
\email{Kenneth.Baldwin@anu.edu.au}
\affiliation{Research School of Physics and Engineering, Australian National University,\\ Canberra, Australian Capital Territory 0200, Australia}
\author{Li-Yan Tang}
\affiliation{Wuhan Institute of Physics and Mathematics, Chinese Academy of Sciences,\\ Wuhan 430071, People’s Republic of China}
\author{A. G. Truscott}
\affiliation{Research School of Physics and Engineering, Australian National University,\\ Canberra, Australian Capital Territory 0200, Australia}

\date{25 May 2015}

\begin{abstract}
We present the first measurement for helium atoms of the tune-out wavelength at which the atomic polarizability vanishes. We utilise a novel, highly sensitive technique for precisely measuring the effect of variations in the trapping potential of confined metastable ($2^{3\!}S_{1}$) helium atoms illuminated by a perturbing laser light field. The measured tune-out wavelength of 413.0938($9_{Stat.}$)($20_{Syst.}$)~nm compares well with the value predicted by a theoretical calculation (413.02(9)~nm) which is sensitive to finite nuclear mass, relativistic, and quantum electro-dynamic (QED) effects. This provides motivation for more detailed theoretical investigations to test QED.
\end{abstract}

\pacs{31.15.ap, 37.10.Vz, 32.10.Dk, 03.75.Kk}

\keywords{Polarizability, Tune Out, Tune-Out, He*, Metastable Helium, Atom Laser, QED, Force Sensing, Optical Dipole Force, Meteorology, Force, Experimental, }
\maketitle



Quantum electrodynamics (QED) is one of the most stringently tested theories in modern physics, and particular interest has focused on QED calculations for helium, the simplest multi-electron atom. Measurements \cite{Zelevinsky2005, Giusfredi2005, Borbely2009} of the transition fine structure intervals for the $2^{3\!}P$ manifold have yielded a test of QED predictions \cite{Pachucki2009} at the one part in $10^{11}$ level with differences of several standard deviations. 
  
Of much lower precision are the experimental and theoretical determinations of transition rates, which are both inherently difficult to measure and predict respectively. Nevertheless, theory and experiment appear to be in good agreement within the (typically of order a few per cent) uncertainty. In helium, we have previously verified theoretical QED predictions in a series of measurements of the transition rates to the ground state for the $2^{3\!}P$ manifold \cite{Dall2008,Hodgman2009a} and the $2^{3\!}S_{1}$ metastable level \cite{Hodgman2009b}.

Recently, QED has been challenged by experiments that determine the proton radius \textit{via} spectroscopy of muonic hydrogen \cite{Pohl2010,Antognini2013}, whose values differ by seven standard deviations (7$\sigma$) from those measured by precision hydrogen spectroscopy (combined with QED theory \cite{Mohr2012}), and by proton-electron scattering experiments \cite{Bernauer2010}. This has created the so-called proton radius puzzle \cite{Pohl2013}. More stringent tests of QED using different experiments are therefore important to provide independent validation or otherwise of QED. 

One such example is the precision measurement of tune-out (or magic-zero \cite{LeBlanc2007}) wavelengths that can provide independent verification of QED predictions for transition rate ratios. At excitation energies above the lowest excited state, the contribution to the dynamic polarizability from the lowest excited state is negative.  There will then occur a series of wavelengths, each associated with a further excited state, where positive contributions to the polarizability from other states will exactly cancel the negative polarizability contributions, thereby creating so-called tune-out wavelengths. 

Mitroy and Tang \cite{Mitroy2013} have estimated theoretically the tune-out wavelengths for transitions from the helium $2^{3\!}S_{1}$ metastable state (He*) to near the $2^{3\!}P$, $3^{3\!}P$ and $4^{3\!}P$ triplet manifolds (at 1083, 389 and 319~nm respectively). These approximate calculations (at around the 0.02\% level) were designed to provide guidance for the first experimental measurements which we present here. Their calculations were based on a composite theory utilizing state-of-the-art transition rate data by Morton and Drake \cite{Drake2007} for the low lying transitions, and model potential oscillator strengths for higher excitations. 

\begin{figure}[h!]
\includegraphics[scale=1.2]{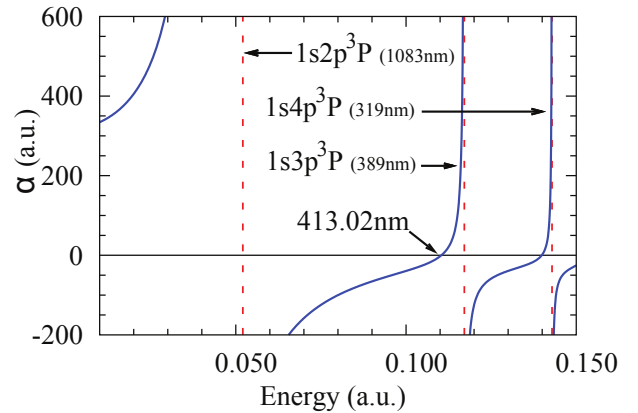}
\caption{\label{fig:epsart} 
(Color online) Helium polarizability spectrum (solid curves) as a function of energy (a.u.). Triplet transition manifold positions are shown by the dotted vertical lines.}
\label{polz}
\end{figure}

From a theoretical perspective, it should be noted that the same QED contributions to the dynamic polarizability are also reflected in the static polarizability. Currently the most accurate theoretical calculation ($<$2~ppm) of the helium ground-state static polarizability \cite{Pachucki2000,Cencek2001,Lach2004} combined with a high-precision experimental measurement (accuracy 9.1 ppm) \cite{Schmidt2007} provides a non-energy test of QED. Our aim is to provide an even more sensitive measurement using the metastable $2^{3\!}S_{1}$ state to yield a non-energy QED test \textit{via} the dynamic polarizability.

Figure \ref{polz} shows the calculated helium dynamic polarizability plotted as a function of energy for He [14]. Tune-out wavelengths are located at the zero crossings between the $2^{3\!}P$, $3^{3\!}P$ and $4^{3\!}P$ manifold transitions (dotted vertical lines), and the key 413 nm tune-out wavelength of interest to this project is marked.

In addition, within each triplet manifold, tune-out wavelengths arise between the spin orbit splittings (not shown). Their positions are determined predominantly by the ratio of oscillator strengths, and therefore are not as sensitive a test of the dynamic polarizability arising from QED \cite{Mitroy2013}. By contrast, the tune-out wavelength at 413~nm (to the long-wavelength side of the 389 nm $3^{3\!}P$ manifold) is expected to be sensitive to finite mass, relativistic, and QED effects upon the transition matrix elements, and its measurement would therefore provide a sensitive test of fundamental atomic structure theory \cite{Mitroy2013}.

Predictions have also been made for tune-out wavelengths in other atomic species \cite{Arora2011,Jiang2013,Cheng2013}. Independent pioneering experiments by Herold \textit{et al.} \cite{Herold2012} and by Holmgren \textit{et al.} \cite{Holmgren2012} yielded the first tune-out wavelength measurements for rubidium and potassium respectively. The accuracy of these experiments improved on the best theoretical values for the transition matrix elements \cite{Arora2011,Mitroy2010}, and provided data that set a limit on the black-body radiation shift in optical clocks. These experiments yielded values for oscillator strength ratios with a precision of a few tenths of a percent, but because of the atomic species used, were not a sensitive test of QED.

Our experiments aim to place QED under further scrutiny by undertaking precision measurements of the He* polarizability to accurately determine the QED-sensitive 413~nm tune-out wavelength. Mitroy and Tang \cite{Mitroy2013} have estimated the location of this tune-out wavelength at the 200~ppm level (413.02(9)~nm). (More recently, another approximate calculation by Notermans \textit{et al.} \cite{Notermans2014} has verified this value within a constant offset that reflects the uncertainty in determining the absolute polarizability.) The dominant uncertainty in the calculated 413~nm tune-out wavelength \cite{Mitroy2013} arises from contributions to the polarizability for transitions to levels of higher energy than the $2^{3\!}P$ and $3^{3\!}P$ manifolds ($\alpha_{remainder}$). Mitroy and Tang further point out that if the 413~nm tune-out wavelength can be determined to an absolute accuracy of 100~fm ($\sim$0.2~ppm), then the fractional uncertainty in the derived atomic structure information would be 1.8 ppm \cite{Mitroy2013}.

Measuring the tune-out wavelength to determine atomic transition rate information has significant advantages over other transition rate measurements. First, the light shift depends only on the transition matrix elements that couple the lower ground or metastable state to higher states, and does not depend on the coupling of those excited states to other states. This contrasts with transition lifetime measurements which necessitate isolating the branching ratios for a range of competing transitions from the excited state to multiple lower states. Second, being a null measurement, the position of the tune-out wavelength (in the weak field limit where we operate) does not depend on knowing the absolute intensity of the light field nor its spatial distribution –- the light intensity simply needs to be stable.

Here we report for the first time the experimental observation of the He* 413~nm tune-out wavelength. This observation was facilitated by the development of a novel, high sensitivity, in-trap cold atom technique based upon a modulated atom laser. We expect that this result will motivate future experiments and theory at even greater accuracy that will further test QED.

\begin{figure}[h]
\includegraphics{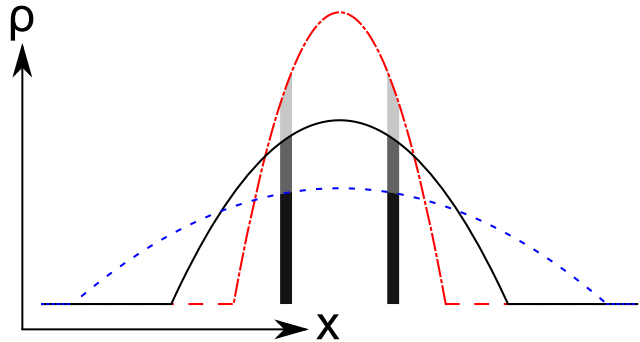}
\caption{\label{fig:epsart}(Color online) Schematic of the He* one-dimensional atomic density profile in the limit of an adiabatically-varied laser potential, with RF out-coupling positions indicated by shaded regions. Black solid line: a purely magnetically trapped BEC. Red dot-dashed line: with an additional attractive laser potential increasing the out-coupling rate. Blue dashed line: with an additional repulsive laser potential decreasing the out-coupling rate.}
\label{exp}
\end{figure}

We undertook experiments to determine the effect of a perturbing light field on a Bose-Einstein Condensate (BEC) of He* atoms using the ultracold atom facility we employed in previous precision spectroscopy measurements  \cite{Dall2008,Hodgman2009a,Hodgman2009b,Vassen2012}. Atoms in the low field seeking $m_{F}=1$ state are confined by a high stability ($\sim$3~nK), asymmetric magnetic trap \cite{Dall2007a,Dedman2007}, yielding trap frequencies of $\sim$~50/500Hz and Thomas-Fermi radii of $\sim$10/100~$\mu m$. We use a linearly polarized ($\frac{1}{2}\sigma^{+} + \frac{1}{2}\sigma^{-}$) light field from a 3~mW tunable diode laser (Moglabs ECD004) and focus it to a  $\sim$20 $\mu m$ diameter spot that overlaps the magnetically trapped atoms. This creates a perturbing potential of $\sim$5~nK ($\sim$1$\times10^{\textrm{-31}}$~J) per nanometre detuning from the tune-out wavelength (Fig. \ref{exp}).

In order to sensitively measure the effect of the introduced optical dipole potential we developed a novel atom-laser-based measurement technique. We continuously output-couple atoms from the magnetic trap using a RF knife \cite{Dall2007} which transfers atoms at a particular magnetic field location in the trap from the low field seeking $m_{F}=1$ state to the magnetically non-interacting $m_{F}=0$ state. The $m_{F}=0$ atoms then free-fall (unperturbed by magnetic fields) onto an 80~mm diameter delay line detector (DLD) located 852 mm below the trap which yields the individual atom arrival time, thereby providing the time-of-flight (TOF) signal. 

When the focused laser beam is on, the additional potential experienced by the atoms alters the atomic density distribution at the RF out-coupling surface and hence the subsequent detection rate (see Fig. \ref{exp}). We then modulate the intensity of the laser beam at a known frequency ($\sim$491~Hz) using an acousto-optic modulator (AOM) and employ Fourier analysis to measure the effect of the light on the output-coupled signal. The frequency of the modulation is set to maximize the detection sensitivity by operating near (but sufficiently well separated from) one of the trapping frequencies ($\sim$500~Hz).

\begin{figure}[h]
\includegraphics[scale=0.55]{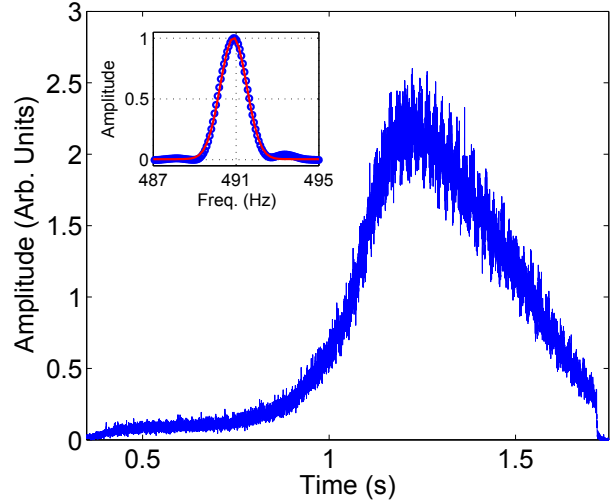}
\caption{\label{fig:epsart}(Color online) Time-of-flight (TOF) DLD signal for the atoms output-coupled from the magnetic trap by the swept RF field in the presence of a modulated perturbing laser field. Inset: Fourier analysis of the TOF signal in the frequency range near the 491 Hz AOM modulation frequency.}
\label{TOF}
\end{figure}

Figure \ref{TOF} shows the DLD signal measured as the RF frequency is swept in order to out-couple a large number of atoms from the optically modulated magnetic trap. The TOF signals were averaged in the time domain and a discrete Fourier transform calculated to isolate the various frequency contributions (such as the 50~Hz AC mains noise visible in Fig. \ref{TOF}). A gaussian was then fit to the spectrum centred at the perturbing laser modulation frequency ($\sim$491 Hz, Fig. \ref{TOF} inset), and the area under the gaussian yielded the perturbation signal amplitude. For each laser wavelength a number of experimental realisations were used, with up to 300 shots at wavelengths with the weakest perturbation signal. To remove any background contributions at this frequency, a TOF-averaged Fourier transform was determined while the probe beam was blocked, and the area thus calculated was subtracted from the signal data over the same frequency range.

The data acquisition itself took many days, so a calibration was used to minimise any systematic drifts. This calibration was performed at 414.00 nm, where the perturbation signal is still appreciable, and was carried out every few hours. The signal was then normalised using the calibration results taken before and after the signal runs. To account for drifts in the probe beam power both the calibration and signal measurements were normalised by the measured average power.

\begin{figure}[h]
\includegraphics[scale=0.45]{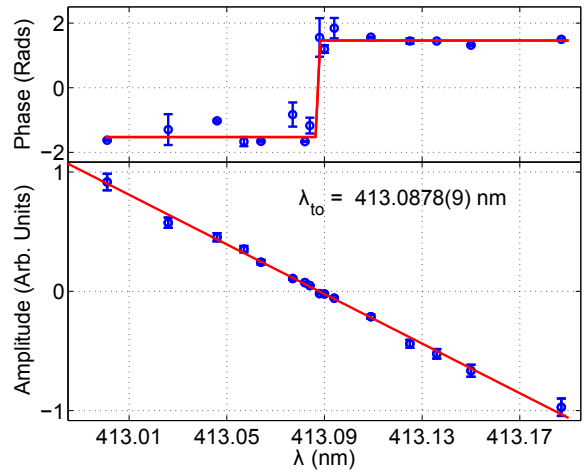}
\caption{\label{fig:epsart}(Color online) Experimental results. Wavelength dependence of the phase (top) and amplitude (bottom) for the modulation signal as a function of wavelength. The solid lines indicate a fit to the data, and the tune-out wavelength thus determined is indicated. }
\label{measurement}
\end{figure}

Figure \ref{measurement} shows the analysed experimental data. Here we plot both the phase and the amplitude of the Fourier-analysed TOF signal in the presence of the modulated perturbing laser light. As can be seen there is a $\pi$ phase shift at the same wavelength as the modulation signal crosses zero. The statistical uncertainty for each point in the amplitude plot represents the $1\sigma$ confidence intervals for the fitted values. Since the amplitude data does not reveal the sign of the perturbation, we use the phase data to fit for the point where the phase crosses from positive to negative. This provides the wavelength at which we can “reflect” the amplitude data – i.e. multiply all data at higher wavelengths by minus one. We then use a weighted fit to a linear function to find the zero of the perturbation signal, thereby yielding the tune-out wavelength.

The raw value for the tune-out wavelength uncorrected for systematic shifts is 413.0878($9_{Stat.}$)~nm, where the 2~ppm uncertainty is statistical. However, there are several potential systematic uncertainties associated with the experimental measurement that need to be addressed in order to compare with QED theory. The wavelength measurement uncertainty at 413~nm arising from the High Finesse WS/7 wavemeter used (0.03~pm or $\sim$0.1~ppm) was negligible, and because we employed ultracold atoms, the effect of Doppler shifts is also negligible ($\sim$0.1~fm). However, systematic uncertainties may arise from the presence of broadband laser light, the presence of the magnetic trapping field, and polarization effects.

The broadband spectrum from the diode laser outside the nominal $\sim$1~MHz linewidth was measured using a high resolution spectrometer and contains $\sim$1$\%$ of the total laser power over a 0.5~nm bandwidth. Following a similar treatment to Holmgren \textit{et al.} \cite{Holmgren2012} we find that the small measured spectral asymmetry around the peak laser wavelength will cause a $-6\pm2$~pm shift in the tune-out wavelength. When we subtract this shift this yields a corrected tune-out wavelength of 413.0938($9_{Stat.}$)($20_{Syst.}$)~nm.

We estimate the remaining sources of systematic shifts to be significantly smaller.  The Zeeman shift at the centre of the magnetic trap is 1.1~MHz. This provides an upper bound of $\sim$1~fm for the effect of magnetic fields on the tune-out wavelength.

In order to estimate the contribution of atoms in different magnetic sub-states, we consider the motion of atoms in the $m_{F}=1$ state whose spin has been flipped by the RF out-coupling field into the $m_{F}=0$ state. These atoms leave the trap either by falling under gravity or by expulsion due to mean field effects. Given the modulation period of the perturbing laser field ($\sim$2~ms) and its beam waist radius ($\sim$10~$\mu m$), $m_{F}=0$ atoms will have left the light field in much less than one modulation period, and will therefore not contribute to the modulation signal. 

Alternatively, atoms may be transferred back to the $m_{F}=1$ state wherein they are recaptured. However, the fractional population that is resonant with the RF out-coupling at a given time is $<1\%$. In addition, based on the theoretical framework \cite{Mitroy2013}, we calculate that this small fraction of $m_{F}=1$ atoms will have a tune-out wavelength that is shifted by $\sim$2~pm from the $m_{F}=0$ tune-out wavelength, with a negligible effect on the measured value.

To account for polarization effects using the theoretical framework \cite{Mitroy2013} we find that for the trapped $m_{F}=1$ atoms, the maximum difference in the tune-out wavelength between pure $\sigma^{+}$ or pure $\sigma^{-}$ polarized light is 3~pm.  However, we have measured the light polarization to be linear within a few per cent (limited by the birefringence of vacuum windows). This yields an uncertainty arising from polarization effects of $\sim$30~fm which is much less than the statistical uncertainty.

The best estimate for the experimental value for the tune-out wavelength is thus 413.0938($9_{Stat.}$)($20_{Syst.}$)~nm, in general agreement with the theoretical estimate of 413.02(9)~nm \cite{Mitroy2013}.  However, the experimental precision (statistical uncertainty) is two orders of magnitude smaller than the uncertainty in the theoretical value. This serves not only to emphasise the sensitivity of this new technique for measuring optical perturbations in the trapping potential, but it also serves as a motivation for improving the theoretical determination of the tune-out wavelength.

To quantify the experimental sensitivity we compared our measured modulation signal response as a function of wavelength, with the theoretical value \cite{Mitroy2013} for the polarizability gradient  at the tune-out wavelength. This yields a polarizability sensitivity of $1.7\times10^{\textrm{-3}}$~a.u. ($2.8\times10^{\textrm{-44}}$~C\,m\textsuperscript{2}\,V\textsuperscript{-1}) at a signal-to-noise ratio of unity. This represents orders of magnitude improvement for the 
polarization sensitivity compared with previous approaches \cite{Herold2012,Holmgren2012} when the lower $d \alpha / d \lambda$ for He* (c.f. Rb and K) is considered.

In conclusion, this first observation of the 413~nm He* tune-out wavelength verifies the theoretical prediction of Mitroy and Tang \cite{Mitroy2013}. The uncertainty level of our novel high-sensitivity light modulation technique for measuring the tune-out wavelength (5~ppm systematic, 2~ppm statistical) has good prospects for significant improvement by simply increasing the laser power and spectral purity, and through a range of other experimental improvements (including studies of the systematics outlined above). There is also the option of using an optical frequency comb to virtually eliminate any uncertainty in the wavelength measurement. This will enable an experimental accuracy approaching 100~fm which would yield a fractional uncertainty in the derived atomic structure information of 1.8~ppm, which compares favourably with static polarizability experiments (9.1~ppm). Finally, we suggest that a significant improvement in the theoretical calculation of the 413~nm He* tune-out wavelength is needed in order to motivate a future experimental campaign to seriously test QED.

We acknowledge many insightful discussions with the late James Mitroy. This work was supported in part by an Australian Research Council Future Fellowship No. FT100100468 held by A.G. Truscott. Li-Yan Tang was supported by the National Basic Research Program of China under Grant No. 2012CB821305, and the NNSF of China under Grants No. 11474319.

\bibliographystyle{apsrev4-1}
\bibliography{Tune-Out}

\end{document}